# Thickness-Dependent Magnetoelasticity and its Effects on Perpendicular Magnetic Anisotropy in Ta|CoFeB|MgOThin Films

P.G. Gowtham[1], G.M. Stiehl[1], D.C. Ralph,[1,2] and R.A. Buhrman[1]

[1]Cornell University, Ithaca, New York, 14853, USA

[2]Kavli Institute at Cornell, Ithaca, New York, 14853, USA

## Abstract

We report measurements of the in-plane magnetoelastic coupling in ultra-thin Ta|CoFeB|MgO layers as a function of uniaxial strain, conducted using a four-point bending apparatus. For annealed samples, we observe a strong dependence on the thickness of the CoFeB layer in the range 1.3-2.0 nm, which can be modeled as arising from a combination of effective surface and volume contributions to the magnetoelastic coupling. We point out that if similar thickness dependence exists for magnetoelastic coupling in response to biaxial strain, then the standard Néel model for the magnetic anisotropy energy acquires a term inversely proportional to the magnetic layer thickness. This contribution can significantly change the overall magnetic anisotropy, and provides a natural explanation for the strongly nonlinear dependence of magnetic anisotropy energy on magnetic layer thickness that is commonly observed for ultrathin annealed CoFeB|MgO films with perpendicular magnetic anisotropy.

Ultra-thin CoFeB|MgO films can possess strong perpendicular magnetic anisotropy (PMA). This observation is of great interest for non-volatile magnetic memory technologies because PMA is required for achieving thermal stability and low write currents at high densities[1]. The total effective anisotropy energy per unit area $K_{eff}t_{eff}$ is commonly analyzed using the Néel model[2], including surface ($K_s$) and volume ($K_V$) contributions to the magnetic anisotropy together with demagnetization effects:

$$K_{eff}t_{eff} = K_s - \left(2\pi M_s^2 - K_V\right)t_{eff}. \qquad (1)$$

Here $t_{eff}$ is the effective thickness of the magnetic layer excluding any dead layer, and $K_{eff}t_{eff} > 0$ corresponds to PMA. However, this simple form generally provides a poor description for the measured thickness dependence of magnetic anisotropy in ultra-thin CoFeB|MgO films possessing PMA. Whereas Eq. (1) predicts a simple linear increase in $K_{eff}t_{eff}$ as a function of decreasing $t_{eff}$ (when $2\pi M_s^2 - K_V > 0$), the measured dependence in films with PMA is often strongly nonlinear, with $K_{eff}t_{eff}$ exhibiting a maximum as a function of decreasing $t_{eff}$ and with the PMA eventually being lost for $t_{eff}$ sufficiently small[3–9]. See, e.g., the data corresponding to the annealed sample in Fig. 1. Non-idealities such as Ta diffusion to the CoFeB|MgO interface during annealing[5,6] are possible reasons for this behavior, but here we suggest that a thickness-dependent magnetoelastic coupling can contribute significantly to this nonlinear $K_{eff}t_{eff}$ vs. $t_{eff}$ dependence observed for ultra-thin CoFeB|MgO with PMA, and may be the dominant explanation for the non-linear behavior.

We investigated Ta(6 nm)/Co$_{40}$Fe$_{40}$B$_{20}$($t_{CoFeB}$)/MgO(2.2 nm)/Hf(1 nm) multilayers deposited by magnetron sputtering onto 375 $\mu m$-thick Si wafers with 500 nm of thermal oxide, and with $t_{CoFeB}$ ranging from 0.7 to 2.0 nm. Details of our film growth are provided in the Supplementary Material (SM). One set of wafers were used for magnetometry. Another set of samples were patterned into 20 μm ×100 μm microstrips using a series of photolithography and subsequent Ar ion milling steps. The microstrips were used for determining magnetoelastic couplings in the Ta|CoFeB|MgO|Hf system. We annealed a full thickness series of both magnetometry samples and microstrip samples at $T = 300$ °C for 1 hour in an in-plane field of 1.3 kOe and at a vacuum pressure of $< 5\times10^{-7}$ torr. The field anneal direction was along the current flow direction of the devices.

Magnetometry measurements on our as-deposited and annealed films were conducted at room temperature using a SQUID magnetometer. The magnetic moments per unit area $M_{sheet}$ are plotted versus $t_{CoFeB}$ in Fig. 2a. For both as-deposited and annealed samples, linear fits of $M_{sheet}$ vs. $t_{CoFeB}$ extrapolate to zero near $t_{CoFeB} = 0$, indicating a negligible magnetic dead layer thickness ($t_{eff} = t_{CoFeB}$). Previous studies of Ta|CoFeB samples have differed regarding the extent of any magnetic dead layers, with some indicating the existence of a dead layer[6,10] as thick as 0.5 nm after annealing and others reporting no dead layer[1,4,11]. Such variation suggests that the extent of any dead layer can depend on the precise choice of sputtering conditions, stack order, processing protocols, and base layer structure[11]. We note that cross-sectional scanning transmission electron microscopy of one of our multilayer samples indicates that the 6 nm Ta base layer is polycrystalline, while other work that found a 0.5 nm dead layer reported that the thinner, ~ 1 nm, Ta base used there appears to be amorphous. From the slopes of the fits in Fig. 2a, the CoFeB saturation magnetization is $M_s = 1120$ emu/cm$^3$ for the as-deposited samples and $M_s$ =1380 emu/cm$^3$ after annealing. The rise in $M_s$ is consistent with B segregation from the CoFe and partial crystallization of the CoFe layer during the annealing.

We characterized the effective anisotropy energy per unit volume $K_{eff}$ by measuring the magnetization $M$ as a function of applied magnetic field $H$ both in ($\parallel$) and out of ($\perp$) the sample plane for samples of each $t_{eff}$ value (e.g., Fig. 2b), and extracted $K_{eff}$ using the expression[12]

$$K_{eff} = M_s \left( \int_0^1 H_\perp(m_\perp) dm_\perp - \int_0^1 H_\parallel(m_\parallel) dm_\parallel \right), \quad (2)$$

where $m_\perp$ and $m_\parallel$ are the normalized components of magnetization out-of-plane and in-plane. The results for $K_{eff} t_{eff}$ vs. $t_{eff}$ are shown as symbols in Fig. 1 for both the as-deposited and annealed samples. The as-deposited samples display good agreement with the linear dependence predicted by the simple Néel model (Eq. (1)) with the fit parameters $K_s$ = 0.3 erg/cm$^2$ and $K_V$ = 1.6 × 10$^6$ erg/cm$^3$. However $K_{eff} < 0$ for all thicknesses (0.7 nm to 2 nm) of the as-deposited samples so that PMA is never achieved. After annealing, PMA is obtained in the thickness range 0.7 nm < $t_{eff}$ < 1.2 nm, but as noted above this generation of PMA occurs simultaneously with a strongly nonlinear $K_{eff} t_{eff}$ vs. $t_{eff}$ behavior in this $t_{eff}$ range. The maximum value of anisotropy per unit area obtained was $K_{eff} t_{eff}$ = 0.26 erg/cm$^2$ at $t_{eff}$ = 0.9 nm. A fit to Eq. (1) for the annealed samples in the range $t_{eff}$ = 1.1 nm to 2.0 nm where $K_{eff} t_{eff}$ vs. $t_{eff}$ is approximately linear yields $K_s$ = 1.5 erg/cm$^2$ and $K_V$ = 0.7 × 10$^6$ erg/cm$^3$.

Following this basic magnetic characterization, we measured the magnetoelastic coupling for the films with in-plane anisotropy using a four-point bend (4PB) strain tester (Fig. 3a). This geometry applies uniaxial strain on the top surface of the substrate, $\delta\varepsilon_{xx}^{top}$, that is uniform between the two inner loading pins and is completely determined by the spacing of the four loading points and the thickness of the substrate:[13,14]

$$\delta\varepsilon_{xx}^{top} = t_s \left(\delta h_{load}\right) / \left(\frac{2}{3}s_2^2 + s_1 s_2\right). \tag{3}$$

The quantity $\delta h_{load}$ is the vertical displacement for the inner pins (a negative number in our experiment) and the other quantities are as defined in Fig. 3a. Here $\varepsilon_{xx} < 0$ corresponds to compression. We use the notation $\delta\varepsilon_{xx}$ rather than simply $\varepsilon_{xx}$ because the CoFeB in Ta|CoFeB|MgO multilayer is, as indicated by $K_V$ and as discussed below, strained even when $\delta h_{load} = 0$. For our geometry, $s_1 = 14$ mm, $s_2 = 8$ mm, and $t_s = 375$ μm. In the limit that the film stack thickness is much smaller than the substrate thickness and the bending is elastic, the mechanically applied strain in the CoFeB can be assumed to be the same as that at the top surface of the Si chip.

Our procedure for measuring the magnetoelastic coupling was to use a lock-in amplifier and Wheatstone bridge to perform a 2 point measurement of the anisotropic magnetoresistance (AMR) as a function of swept magnetic field for different fixed values of applied bending strain, using 5.3 mm × 45 mm device dies with the long axis along the x direction in Fig. 3b, perpendicular to the current direction in the sample. Each die was loaded with the long axis bridging the support pins of the 4PB apparatus, inside an electromagnet capable of applying a magnetic field in the x direction. For the annealed in-plane magnetized samples ($t_{eff} > 1.2$ nm), the magnetic easy axis was set along the y-direction by the field orientation during annealing. For the as-deposited samples the y-axis was made the easy axis by applying a sufficient compressive strain in the x-direction (usually $|\delta\varepsilon_{xx}| > 0.05\%$). For both types of samples, therefore, the applied magnetic field was oriented along the in-plane hard axis, and produced a non-hysteretic rotation of the magnetization. Under these conditions, the AMR curve serves as a faithful representation of the in-plane rotation angle $\varphi$ for the average magnetization, with $R(\varphi) = R_0 + \Delta R \sin^2\varphi$. The resistance is a minimum when the magnetization is saturated along the x-axis.

Because of the magnetoelastic coupling, the application of a compressive bend strain alters the shape of the AMR curve by changing the in-plane magnetic anisotropy energy. Figure 3c shows the measured AMR for a $t_{eff} = 1.7$ nm sample with $\delta\varepsilon_{xx}$ ranging from 0 to -0.095% in increments of -0.0065%. The AMR curves are normalized for each value of strain by mapping the (large-field) saturated resistance value to 0 and the (zero-field) maximum resistance value to 1. With the identification that the normalized x-component of magnetization is $m_x = \cos[\arcsin[\sqrt{R_{norm}}]]$, the AMR measurements can then be transformed to yield $H_x(m_x)$ vs. $m_x$ curves (shown in Fig 3d), from which the in-plane uniaxial magnetoelastic coupling $B_{eff}^{uniaxial}$ multiplied by the strain can be calculated as:[15]

$$(\delta\varepsilon_{xx}) B_{eff}^{uniaxial} = \frac{\frac{3}{4} M_s \left( \int_{m_1}^{m_2} H_x(m_x, \delta\varepsilon_{xx}) dm_x - \int_{m_1}^{m_2} H_x(m_x, 0) dm_x \right)}{(m_2^2 - m_1^2)} \equiv \Lambda. \quad (4)$$

Here $m_1$ and $m_2$ are normalized magnetization points in the $H_x$ vs. $m_x$ curve that are used as limits of integration (we use $m_1 = 0.4$ and $m_2 = 0.8$). Equation (4) holds under the assumption that the Poisson ratio $\nu = 1/3$ (previous experiments report $0.25 < \nu < 0.4$ for metals in thin film form[16–18] and in bulk[19,20]) and that the CoFeB is an isotropic medium in the plane of the film.

The latter assumption should be accurate for our samples as no in-plane texturing is expected in annealed Ta/CoFeB/MgO multilayers, although there is strong grain-by-grain out-of-plane texturing at the CoFeB|MgO interface.[21] We determine $B_{eff}^{uniaxial}$ by fitting $\Lambda$ (the right hand side of Eq. (4)) vs. $\delta\varepsilon_{xx}$ to a straight line for each sample and evaluating the slope as shown in the inset of Fig 3d.

Measurements of the in-plane magnetoelastic coupling for the samples in which the magnetic anisotropy is in plane are shown in Fig. 4 for both the as-deposited and annealed Ta|CoFeB|MgO thin films. For the as-deposited samples, $B_{eff}^{uniaxial}$ is approximately constant, near $-4 \times 10^7$ erg/cm$^3$ for all values of $t_{eff}$. As will be discussed below, if we assume that the volume anisotropy of the as-deposited sample, $K_V = 1.6 \times 10^6$ erg/cm$^3$, arises only from a biaxial elastic strain $K_V = B_{eff}^{biaxial} e_{biaxial}^{as-deposited}$ (where $B_{eff}^{biaxial}$ is the component of the magnetoelastic tensor coupling to biaxial strains), the measured uniaxial magnetoelastic constant $B_{eff}^{uniaxial} \approx -4 \times 10^7$ erg/cm$^3$ suggests a large, and an approximately thickness-independent, biaxial compressive strain $e_{biaxial}^{as-deposited}$ in the as-deposited samples. To estimate this strain we assume that

$$B_{eff}^{biaxial} \approx B_{eff}^{uniaxial} + B_{eff}^{13}, \qquad (5)$$

where $B_{eff}^{13}$ is the term that connects the magnetic free energy to strains perpendicular to the sample plane.[15] This relationship is appropriate for the condition of isotropy in the sample plane and a Poisson ratio $\nu \approx 1/3$, within the typical range found for metals. For purposes of estimation here we can also assume $B_{eff}^{13} \approx B_{eff}^{uniaxial}$, which is appropriate for an isotropic system, such as the amorphous as-deposited film. This analysis indicates that $e_{biaxial}^{as-deposited} \approx -0.02$, which is consistent with the strong compressive strain that is common in sputter-deposited refractory metal films.[22]

After annealing, our results show that, while the volume anisotropy is substantially reduced, the magnitude of $B_{eff}^{uniaxial}$ is considerably larger and has a pronounced dependence on $t_{eff}$, changing by 60 % between $t_{eff}$ = 1.3 and 2.0 nm. This fits well to the functional form $B_{eff}^{uniaxial} = (B_s^u / t_{eff}) + B_V^u$, suggesting that in the annealed samples there is both a strong volume magnetoelastic coupling $B_V^u$ = -1.5×10$^8$ erg/cm$^3$ and strong effective interfacial magnetoelastic coupling $B_s^u$ = +12.1 erg/cm$^2$. One possible mechanism for an apparent interfacial magnetoelastic coupling term is a large second-order term $D$ in the volume magnetoelastic coupling, $B_{eff}^{uniaxial} = B_V^0 + D\varepsilon$, in combination with a strongly thickness-dependent strain in the annealed samples, as observed in coherent, epitaxial bilayers with a strong crystalline mismatch[22]. Alternatively or in addition, another possible mechanism for the observed thickness dependence is that the same interface electronic effect (i.e. Fe-3d/O 2p hybridization)[24,25] at the CoFeB|MgO interface responsible for the strong PMA in the thin annealed and partially crystallized CoFe(B) films[26–32] also contributes an interface-like term to the effective magnetoelasticity of the annealed samples.

If the strain in annealed CoFeB|MgO samples is thickness-dependent in the presence of a strongly thickness-dependent magnetoelasticity, the magnetic anisotropy will be strongly altered. To analyze this effect, we employ a generalization of the Néel model that explicitly takes into account the magnetoelastic contribution to the magnetic energy[32,33]:

$$K_{eff} t_{eff} = K_s^0 + \left[ K_V^0 - \left(2\pi M_s^2\right) + B_{eff}^{biaxial}\left(t_{eff}\right) \varepsilon_{biaxial}\left(t_{eff}\right) \right] t_{eff}. \quad (6)$$

Here $K_s^0$ and $K_V^0$ are the surface and volume magnetoelastic couplings at zero strain. We consider the case relevant to CoFeB|MgO films without in-plane texture, where the strain associated with growth and annealing should be biaxial, and for simplicity we assume that the average strain variation with CoFeB thickness can be approximated as

$$\varepsilon_{biaxial}\left(t_{eff}\right) \approx \varepsilon_0^{biaxial} + \left(\gamma_{biaxial} / t_{eff}\right) \quad (7)$$

over the thickness range $t_{CoFeB} = 0.7 - 2.0$ nm employed in our study. The precise functional form is not essential for our conclusions (see the SM for further discussion of this point). The expression we have chosen to model the variation in strain has the virtue of yielding a particularly simple extension of the Néel form for $K_{eff} t_{eff}$. Generally, the effect of the thickness-dependent magnetoelasticity on the magnetic anisotropy requires that the strain in the CoFeB layer increases strongly after annealing with decreasing $t_{CoFeB}$ for the thinner films in the range studied, and varies much less strongly or not at all for the samples with $t_{CoFeB} > 1.2$ nm. We note that $B_{eff}^{biaxial}$ (Eq. (5)) involves a different combination of magnetoelastic tensor elements than $B_{eff}^{uniaxial}$. Neither $B_{eff}^{13}$ nor $B_{eff}^{biaxial}$ have been measured for Ta|CoFeB|MgO samples, either as-deposited or annealed. However, as long as the overall magnetoelastic coupling $B_{eff}^{biaxial}$ has a significant interface term, with $B_{eff}^{biaxial} \approx (B_s^b / t_{eff}) + B_V^b$, then it follows from Eq. (6) that the total magnetic anisotropy per unit area should approximately possess a simple, separable functional form

$$K_{eff} t_{eff} = (K_V^f - 2\pi M_s^2) t_{eff} + K_s^f + \frac{K_3}{t_{eff}}, \tag{8}$$

containing an effective volume term with coefficient $K_V^f = K_V^0 + B_V^b \varepsilon_0^{biaxial}$, an effective surface term with coefficient $K_s^f = K_s^0 + B_V^b g_{biaxial} + B_s^b e_0^{biaxial}$, and a term scaling as $t_{eff}^{-1}$ with coefficient $K_3 = B_s^b \gamma_{biaxial}$. The dashed line in Fig. 1 is a 3-parameter fit of Eq. (7) to the data for our annealed Ta|CoFeB|MgO samples, with $K_V^f - 2\pi M_s^2 = (-1.77 \pm 0.03) \times 10^7$ ergs/cm³, $K_s^f = +3.25 \pm 0.03$ ergs/cm², and $K_3 = (-1.28 \pm 0.03) \times 10^{-7}$ erg/cm.

An accurate quantitative analysis of the different contributions to the anisotropy requires knowing the value of $B_{eff}^{biaxial}$ or equivalently both $B_{eff}^{uniaxial}$ and $B_{eff}^{13}$. In principle, $B_{eff}^{biaxial}$ can be measured by a biaxial strain test (e.g. a ring-on-ring test). However, for purposes of estimation here we will assume $B_{eff}^{13} \approx B_{eff}^{uniaxial}$, so that $B_{eff}^{biaxial}(t_{eff}) \approx 2 B_{eff}^{uniaxial}(t_{eff})$. This assumption is rigorous in systems with full isotropy or cubic symmetry, but will not be rigorous for our samples due to out-of-plane texturing of the CoFeB film and symmetry breaking at the CoFeB|MgO interface. Given our determination that $B_s^u = +12.1$ erg/cm² provides a good fit to the measured $t_{eff}$ dependence of $B_{eff}^{uniaxial}$ of the annealed samples (Fig. 4), an explanation of the nonlinearity in $K_{eff} t_{eff}$ versus $t_{eff}$ entirely in terms of thickness-dependent magnetoelastic coupling then requires that $\gamma_{biaxial}$ have the value $\gamma_{biaxial} \approx K_3/(2B_s^u) = -0.053 \pm .002$ nm, and that $e_0^{biaxial} \approx K_V^f/(2B_V^u) = 0.019$, under the assumption $K_V^0 = 0$. The negative sign of $\gamma_{biaxial}$ here corresponds to a greater magnitude of compressive strain for thinner CoFeB films and a lower compressive strain ($e_{biaxial} \leq -0.016$) as the CoFeB film gets into a higher thickness range 1.5 nm $< t_{eff} < 2$ nm. The magnitude that we estimate for $\gamma_{biaxial}$ corresponds to a total change in average CoFeB film strain over the thickness range of our anisotropy measurements (0.7 – 2.0 nm) of $\Delta e^{biaxial} = 0.05$.

The presence of compressive strains in the CoFeB film estimated by the preceding analysis conflicts with strains predicted from consideration of the equilibrium lattice mismatch between hetero-epitaxial thin film layers of MgO and CoFe at a coherent interface. If we assume bulk equilibrium lattice constants, the CoFeB (MgO) should be under tensile (compressive) rather than the compressive (tensile) stress[6,35]. However the available experimental evidence agrees with our conclusions regarding the presence of compressive strain. Recent X-ray diffraction measurements on CoFeB(6nm)/MgO(2nm) multilayers have reported that the equilibrium lattice spacing of bulk MgO is not observed in these layers, but rather that the MgO lattice is considerably expanded and thus is under tensile stress, with this expansion decreasing with higher annealing temperatures (250 °C to 400 °C). The tensile strain can potentially be attributed to point defects, which are generally found to expand the lattice of non-stoichiometric oxides.[36,37] The X-ray work also reports that the annealing process results in the formation of textured "nanopipes"[21] as crystalline CoFe grains nucleate at the CoFeB|MgO interface. The process occurs through crystallization templating of the CoFeB off the MgO surface and is accomplished by B out-diffusion. The study shows that these nano-columnar, partially crystallized CoFe grains are under high compressive biaxial strain.[38] The CoFe lattice parameter reported is compressed more than 3% below the bulk equilibrium value for a 300 °C annealed multilayer, resulting in an average lattice parameter difference between the MgO and CoFe greater than 7% (4.5%) for a 300 °C (400 °C) CoFeB(6 nm)|MgO(2 nm) sample.

We surmise therefore that coherent heteroepitaxy is not the dominant factor in determining the strain configuration adjacent to these thin CoFe|MgO bi-layers as it cannot be responsible for the compressive strain on the CoFe grains as observed by this X-ray study, and as needed to account for the thickness-dependent magnetic anisotropy in terms of a interfacial magnetoelastic effect as we are proposing here. We speculate that the CoFe compressive strain arises instead from the B displacement during the nucleation and growth of the textured CoFe|MgO nanopipes that are required for high TMR, and presumably for strong interfacial anisotropy.

In summary, we have measured the magnetoelastic coupling in annealed Ta|CoFeB|MgO samples for in-plane uniaxial strains, finding a strong dependence on the thickness of the CoFeB layer that can be modeled in terms of large volume and surface contributions to the magnetoelastic coupling. We suggest that a thickness-dependent magnetoelastic coupling and thickness-dependent elastic strain can together have a significant influence on the strength and thickness dependence of PMA in annealed Ta|CoFeB|MgO samples. In particular, thickness-dependent magnetoelastic coupling provides a natural explanation for the functional form of the nonlinearity commonly observed for thin magnetic layers with PMA in their curves of $K_{eff}t_{eff}$ vs. $t_{eff}$. More detailed measurements of the biaxial magnetoelastic coupling and characterization of the strain distribution in ultrathin CoFeB|MgO bilayers are thus warranted. A clear understanding of the strain distribution in nanocolumnar CoFeB formed by templating off of nanocrystalline MgO, and the role that B diffusion and various NM underlayers are playing in this distribution are currently lacking. A clear picture of the interplay between the biaxial magnetoelastic coupling, strain distribution, and materials physics/chemistry in NM|CoFeB|MgO systems would allow for deeper insight into the behavior of the PMA in these systems, and potentially offers new routes for tailoring the PMA for technological applications.

**Acknowledgements**

We thank R.B. van Dover for useful discussions and Nathan Ellis for aiding us in the design of the 4PB setup. We also thank G.E. Rowlands, L.H. Villela-Leao, J. Park, Y.X. Ou, and S.V. Chakram for assistance. This research was supported by ONR, ARO, and NSF (DMR-1406333), and made use of the Cornell Center for Materials Research Shared Facilities which are supported through the NSF MRSEC program (DMR-1120296). This work was also performed in part at the Cornell NanoScale Facility, a member of the National Nanotechnology Infrastructure Network, which is supported by the National Science Foundation (Grant ECCS-15420819).GMS acknowledges support by a National Science Foundation Graduate Research Fellowship under Grant No. DGE-1144153.

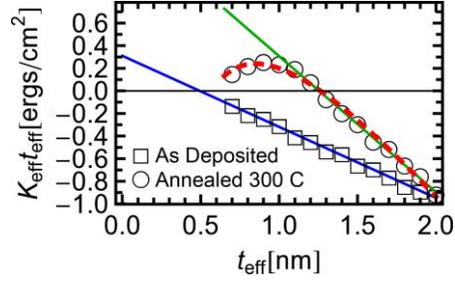

**Figure 1.** $K_{eff}t_{eff}$ vs. $t_{eff}$ for the as-deposited and annealed Ta(6 nm)|Co$_{40}$Fe$_{40}$B$_{20}$($t_{CoFeB}$)|MgO(2.2 nm)|Hf(1 nm) thickness series. The $K_{eff}t_{eff}$ data for the as-deposited samples fit well to a Néel model with $K_s \sim 0.3$ ergs/cm$^2$. The annealed data are compared to a Néel model fit (solid green) and to a model including thickness-dependent magnetoelastic interactions (dashed red).

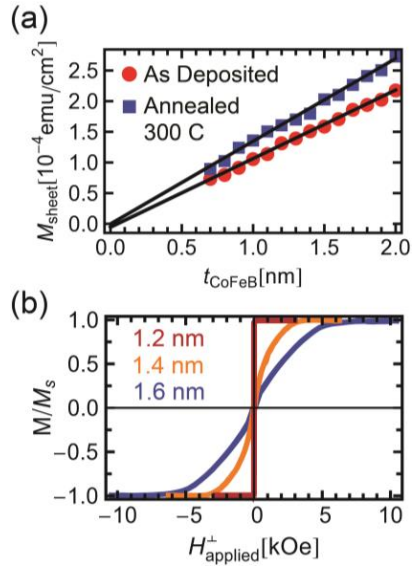

**Figure 2. a)** The magnetic moment sheet density $M_{sheet}$ vs. nominal CoFeB film thickness for the as-deposited and annealed films. The slopes of linear fits to the data yield $M_s = 1120$ emu/cm$^3$ and $M_s = 1380$ emu/cm$^3$ for the as-deposited and annealed samples, respectively. No appreciable magnetic dead layers are found in our samples either as-deposited or after annealing. **b)** SQUID scans of annealed films with field oriented perpendicular to the film plane for $t_{CoFeB} = $ 1.2, 1.4, and 1.6 nm. The transition to PMA occurs near 1.2 nm.

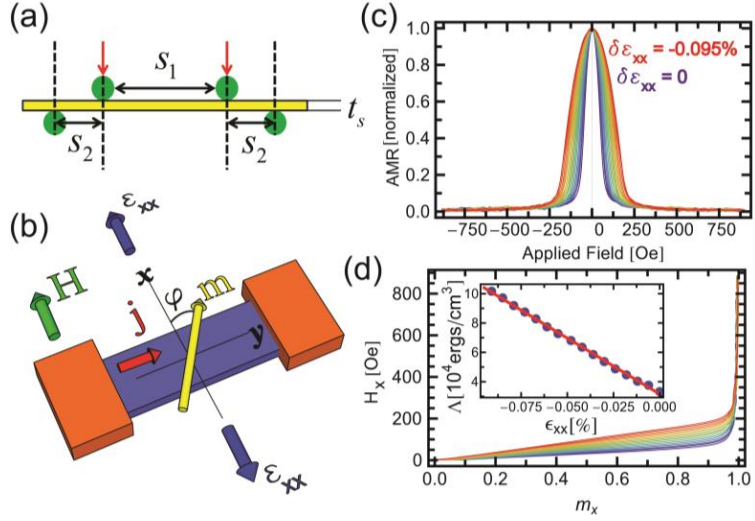

**Figure 3. a)** Schematic of the 4PB setup. **b)** Micro-wire device layout and geometry used for measuring AMR and extracting $B_{eff}^{uniaxial}$. **c)** Normalized MR hard axis curve series for an annealed device with $t_{CoFeB} = 1.7$ nm, as a function of increasing compressive strain. The $\delta\varepsilon_{xx}$ increment between each AMR sweep is $-0.0065\%$. **d)** Conversion of AMR field sweeps to $H_x(m_x)$ curves. [**Inset:** Change in anisotropy energy density as a function of strain. The slope yields $B_{eff}^{uniaxial} \sim -7.6 \times 10^7$ erg cm$^3$ for the annealed sample with $t_{CoFeB} = 1.7$ nm.]

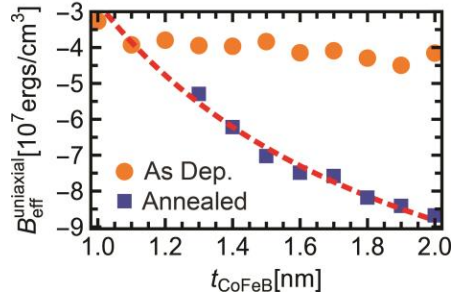

**Figure 4.** $B_{eff}^{uniaxial}$ vs. $t_{CoFeB}$ for the samples as-deposited and annealed at 300 C for 1 hour. The dashed red line is a fit to the $B_{eff}^{uniaxial}$ vs. $t_{CoFeB}$ data for the annealed series using the functional form $B_{eff}^{uniaxial} = (B_s^u / t_{eff}) + B_V^u$. The values $B_s^u = +12.1$ ergs/cm$^2$ and $B_V^u = -1.5 \times 10^8$ ergs/cm$^3$ are extracted from the fit.

# Thickness-Dependent Magnetoelasticity and its Effects on Perpendicular Magnetic Anisotropy in Ta|CoFeB|MgO Thin Films

# SUPPLEMENTARY MATERIAL

## S1. Film Growth and Characterization Details

Our Ta(6 nm)/Co$_{40}$Fe$_{40}$B$_{20}$($t_{CoFeB}$)/MgO(2.2 nm)/Hf(1 nm) multilayers were sputtered in an AJA ATC 2200 Magnetron Sputtering system. The base pressure in our chamber was $P_0 < 1.0 \times 10^{-8}$ torr and the working Ar gas pressure was kept at 2 mtorr throughout the deposition process. All metallic films were DC sputtered at a power of 30 W, resulting in rates of ~0.1 Å/sec for Ta and Hf and ~0.08 Å/sec for CoFeB. MgO was sputtered at 100 W RF at a rate of ~0.04 Å/sec. All film stacks were deposited under stage rotation. Magnetometry was conducted on 3.5 × 3.5 mm$^2$ square dies using a Quantum Design MPMS SQUID magnetometer. The chips were diced using a K&S 7100 Dicing SAW. Wafers used for the 4PB bend tests were diced into 5.3 mm × 45 mm chips using a special-purpose S1235Si blade so as not to damage the sides of the die. This was essential for allowing us to apply significant strain to the chip without shattering it. We were able to generate $\delta\varepsilon_{xx}$ ~ -0.12% corresponding to a chip deflection of $\delta h_{load}$ = -0.5 mm before significant risk of shattering the chip.

## S2. Magnetoelastic Anisotropy Energy Density

There are two independent magnetoelastic couplings $B_{eff}$ and $B_{eff}^{13}$ which govern the connection between the magnetic anisotropy energy, the in-plane strains $\varepsilon_{xx}, \varepsilon_{yy}$, and the strain perpendicular to the film plane $\varepsilon_{zz}$ in out-of plane textured, in-plane-isotropic ultra-thin films.

The magnetic anisotropy energy density arising from the magnetoelastic interaction can be expressed in this case as

$$f_{ME} = B_{eff}(\varepsilon_{xx}\alpha_1^2 + \varepsilon_{yy}\alpha_2^2) + B_{eff}^{13}\varepsilon_{zz}\alpha_3^2, \tag{1}$$

where $\alpha_1$, $\alpha_2$, and $\alpha_3$ are the angular cosines relative to the $x$, $y$, and $z$ axes respectively. For in-plane magnetized samples, $B_{eff}$ can be extracted by applying a uniaxial $\delta\varepsilon_{xx}$ strain and measuring changes in the in-plane anisotropy energy. For this situation $\alpha_3 = 0$ and so contributions arising from the $B_{eff}^{13}$ magnetoelastic coupling rigorously vanish. An applied uniaxial strain $\delta\varepsilon_{xx}$ also will result in Poisson strains in the $y$ and $z$ directions. The strain $\delta\varepsilon_{yy} = -\delta\varepsilon_{xx}/3$ assuming a Poisson ratio $\nu = 0.3$. Using $\alpha_1^2 + \alpha_2^2 = 1$ then simplifies Supplementary Eqn. (1) in the in-plane magnetized situation with a uniaxial strain $\delta\varepsilon_{xx}$ to:

$$f_{ME} = \frac{4}{3}B_{eff}\delta\varepsilon_{xx}\alpha_1^2 = \frac{4}{3}B_{eff}^{uniaxial}\delta\varepsilon_{xx}\alpha_1^2 \tag{2}$$

We have dropped terms that do not depend on the magnetization orientation and have defined the in-plane uniaxial magnetoelastic coupling $B_{eff}^{uniaxial} = B_{eff}$ as used in the main text. Differences in the anisotropy energy density for field sweeps along the in-plane hard axis (assumed to be the $x$ axis) at different $\delta\varepsilon_{xx}$ can then be used to directly extract $B_{eff}^{uniaxial}$. Supplementary Eqn. 2 and a curve integration method on the resultant $m_x$ vs. $H_x$ curves at different strains are easily combined to yield Eqn. 4 of the main text.

We now derive the magnetoelastic anisotropy energy density arising from in-plane biaxial strain $\varepsilon_{xx} = \varepsilon_{yy} = \varepsilon_{biaxial}$. The film will experience a Poisson strain in the $z$ direction $\varepsilon_{zz} = \frac{-2\nu}{1-\nu}\varepsilon_{biaxial}$ which simply reduces to $\varepsilon_{zz} = -\varepsilon_{biaxial}$ with the assumption $\nu = 0.3$. Supplementary Eqn. (1) then becomes $f_{ME} = B_{eff}^{uniaxial}\varepsilon_{biaxial}(\alpha_1^2 + \alpha_2^2) - B_{eff}^{13}\varepsilon_{biaxial}\alpha_3^2$. Using the identity $\alpha_1^2 + \alpha_2^2 + \alpha_3^2 = 1$, we get the form for the magnetoelastic contribution to the magnetic anisotropy energy:

$$f_{ME} = -\left(B_{eff}^{uniaxial} + B_{eff}^{13}\right)\varepsilon_{biaxial}\alpha_3^2 = -B_{eff}^{biaxial}\varepsilon_{biaxial}\alpha_3^2 \qquad (3)$$

Thus in-plane biaxial strains will modify the PMA through the biaxial magnetoelastic coupling $B_{eff}^{biaxial}$ which under the assumption $\nu = 0.3$ is $B_{eff}^{biaxial} = B_{eff}^{uniaxial} + B_{eff}^{13}$.

## S3. Thickness Dependence of the Average Strain

We have fit our $K_{eff}t_{eff}$ data to a form that takes into account volume and effective surface magnetoelastic effects (i.e. $B_{eff}^{biaxial}(t_{eff}) \approx (B_s^b / t_{eff}) + B_V^b$):

$$K_{eff}t_{eff} = K_s^0 + \left[K_V^0 - (2\pi M_s^2) + B_{eff}^{biaxial}(t_{eff})\varepsilon_{biaxial}(t_{eff})\right]t_{eff} \qquad (4)$$

where the values for $B_s^b = +24$ ergs/cm$^2$ and $B_V^b = -3.0 \times 10^8$ ergs/cm$^3$ are same as in the main body of the text and where we have assumed a form $\varepsilon_{biaxial}(t_{eff}) = \gamma_{biax}t^{-\alpha} + \varepsilon_0$, in the region from $t_{eff} = 0.7 - 2.0$ nm. We have investigated different values of $\alpha$ ranging from 0.25 to 1, which corresponds to a large range of power law behavior observed in various magnetron sputtered films. The specific exponent that governs the average strain thickness behavior will depend on the film growth mode and is known to vary with Ar processing gas pressure and sputtering power.[1] Supplementary Fig. 1 shows the results of the fits and Supplementary Table 1 shows the best fit parameters for each value of $\alpha$. Supplementary Fig. 1 implies that the fit behavior to $K_{eff}t_{eff}$ is not extremely sensitive to $\alpha$ or quite generally to the precise functional form of the thickness dependence of the average strain in the film (for the thickness regime we are studying) – provided that the change in the film strain as a function of film thickness is large enough in the ultrathin regime and becomes small enough as the CoFeB is made thicker.

| $\alpha$ | $\gamma_{biaxial}$ | $\varepsilon_0$ | $K_s^0$ |
|---|---|---|---|
| 0.25 | -.115 nm$^{.25}$ | 0.12 | 1.2 ergs/cm$^2$ |
| 0.5 | -0.087 nm$^{.5}$ | 0.054 | 1.2 ergs/cm$^2$ |
| 0.75 | - 0.064 nm$^{.75}$ | 0.031 | 1.2 ergs/cm$^2$ |
| 1.0 | - 0.053 nm | 0.019 | 1.2 ergs/cm$^2$ |

**Supplementary Table 1.** Best fit parameters to the data using the energy density of Eqn. 4 and a strain function $\varepsilon_{biaxial}(t) = \gamma_{biaxial} t^{-\alpha} + \varepsilon_0$.

## Supplementary References

1. Janssen, G. C. A. M. Stress and strain in polycrystalline thin films. *Thin Solid Films* **515,** 6654–6664 (2007).

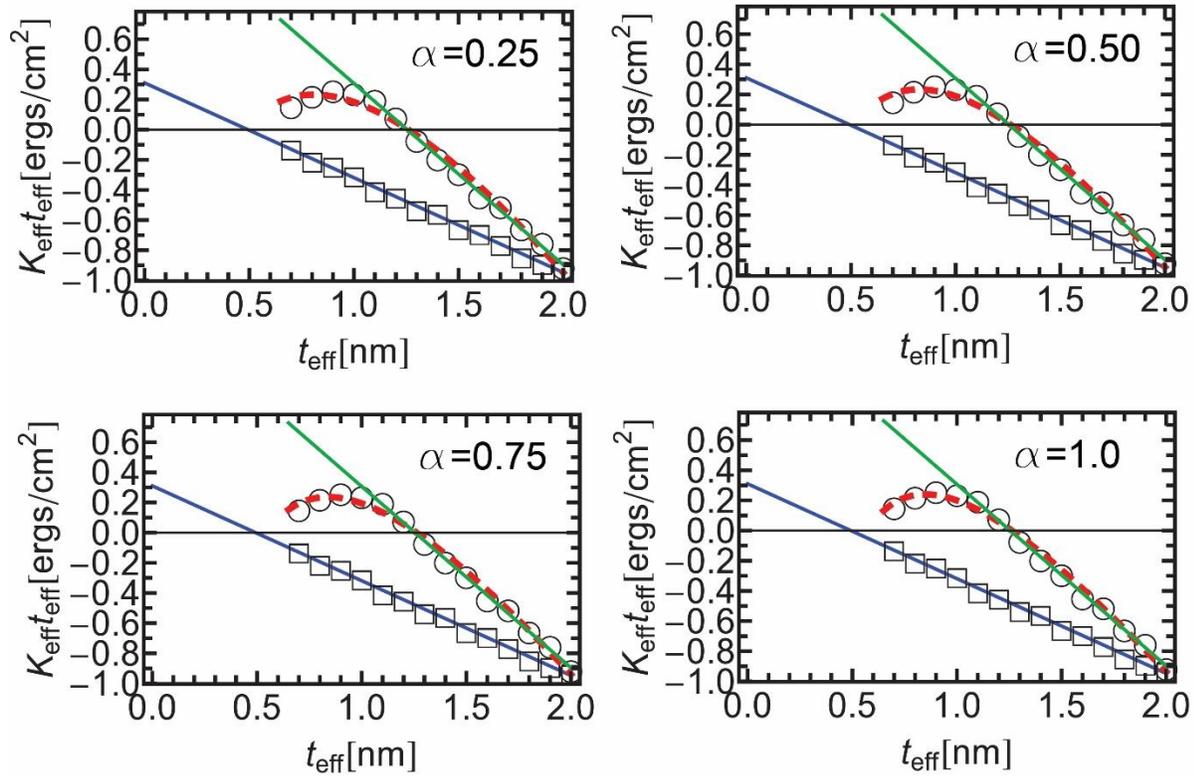

**Supplementary Figure 1.** The red dashed lines correspond to the nonlinear model for $K_{eff}t_{eff}$ including volume and effective surface magnetoelastic couplings with $\varepsilon_{biaxial}(t_{eff}) = \gamma_{biax} t^{-\alpha} + \varepsilon_0$, for $\alpha$ ranging from 0.25 to 1.0.